\documentstyle[12pt]{article}
\def\figcap{\section*{Figure Captions\markboth
        {FIGURECAPTIONS}{FIGURECAPTIONS}}\list
        {Figure \arabic{enumi}:\hfill}{\settowidth\labelwidth{Figure
99:}
        \leftmargin\labelwidth
        \advance\leftmargin\labelsep\usecounter{enumi}}}
 \relax
\def\tabcap{\section*{Tables\markboth
        {TABLES}{TABLES}}\list
        {Table \arabic{enumi}:\hfill}{\settowidth\labelwidth{Table
9:}
        \leftmargin\labelwidth
        \advance\leftmargin\labelsep\usecounter{enumi}}}
 \relax
\newskip\humongous \humongous=0pt plus 1000pt minus 1000pt

\newif\ifdtup

\newcommand{\BLNV}{\mbox{\small{BLNV}}}
\newcommand{\CKM}{\mbox{\small{CKM}}}
\newcommand{\HW}{\mbox{\small{HERWIG}}}
\newcommand{\HERBVI}{\mbox{\small{HERBVI}}}
\newcommand{\MC}{\mbox{Monte Carlo}}
\newcommand{\IPROC}{\mbox{\tt{IPROC}}}
\newcommand{\GENEV}{\mbox{\tt{GENEV}}}
\newcommand{\QCD}{\mbox{\small{QCD}}}
\newcommand{\MAMBO}{\mbox{\small{MAMBO}}}
\newcommand{\HVN}{\mbox{{\small HERBVI} version~1.0}}
\newcommand{\mb}[1]{\ifmmode#1\else\mbox{$#1$}\fi}
\newcommand\al{\mb{\alpha}}
\newcommand\be{\mb{\beta}}
\newcommand\de{\mb{\delta}}
\newcommand\De{\mb{\Delta}}
\newcommand\si{\mb{\sigma}}
\newcommand\ga{\mb{\gamma}}
\newcommand\th{\mb{\theta}}
\newcommand\ze{\mb{\zeta}}
\newcommand{\shat}{\mb{\hat{s}}}
\newcommand{\calP}{\mb{{\cal P}}}
\newcommand{\myfrac}[2]{\mbox{\small$\frac{#1}{#2}$}}
\newcommand{\s}[1]{\mbox{\small#1}}
\def\routine#1{
\noindent{{\tt{SUBROUTINE\ #1}}}
}
\def\function#1{
\noindent{{\tt{FUNCTION\ #1}}}
}
\begin{document}
\bibliographystyle{bibstyle}
\begin{titlepage}
\renewcommand{\thefootnote}{\fnsymbol{footnote}}
\begin{flushright}
     Cavendish--HEP--94/16 \\
     Liverpool--HEP--95/201 \\
     hep-ph/9504232
     \end{flushright}
\vspace*{\fill}
\begin{center}
{\Large \bf HERBVI - a program for simulation of \\
baryon and lepton number violating processes\footnote{
Research supported in part by the UK Science and Engineering Research
Council and by the EC Programme ``Human Capital and Mobility", Network
``Physics at High Energy Colliders", contract CHRX-CT93-0357 (DG 12 COMA).}
}
\end{center}
\par \vskip 2mm
\begin{center}
        {\bf M.J.\ Gibbs\footnote{Current Address: Department of Physics,
Oliver Lodge Laboratory, P.O.~Box~147, Liverpool University,
Liverpool~L69~3BX.}~and B.R.~Webber} \\
        Cavendish Laboratory, University of Cambridge \\
        Madingley Road, Cambridge CB3 0HE, U.K.
        \par \vskip 2mm \noindent
\end{center}
\par \vskip 2mm

\begin{center} {\large \bf Abstract} \end{center}
\begin{quote}

We describe a Monte Carlo event generator for the simulation
of baryon- and lepton-number violating processes at
supercolliders. The package, {\HERBVI}, is designed
as a hard-process generator interfacing to the general
hadronic event simulation program {\HW}. In view of the
very high multiplicity of gauge bosons expected in
such processes, particular attention is paid to
the efficient generation of multiparticle phase space.
The program also takes account of the expected colour
structure of baryon-number violating vertices, which has
important implications for the hadronization of the
final state.

\end{quote}

\begin{center}
{\it Submitted to Comp. Phys. Comm.}
\end{center}

\vspace*{\fill}
\begin{flushleft}
     Cavendish--HEP--94/16 \\
     Liverpool--HEP--95/201 \\
     \today
\end{flushleft}
\end{titlepage}
\thispagestyle{empty}

\section*{Program summary}

\begin{itemize}

\item[] {\em{Title of program: }}{\tt{HERBVI}}

\item[] {\em{Program obtainable from: }}{\tt{gibbs@dxcern.cern.ch}}

\item[] {\em{Prgramming language used: }}{\tt{FORTRAN 77}}

\item[] {\em{Memory required: }}approx. 120 kbytes

\item[] {\em{Number of bits per word: }}32

\item[] {\em{Subprograms used: }}{\tt{HERWIG}} version 5.7 or later

\item[] {\em{Number of lines in distributed program: }}4000

\item[] {\em{Keywords: }}baryon and lepton number violation, Monte Carlo
techniques, high particle multiplicites

\item[] {\em{Nature of physical problem: }}simulation of baryon
and lepton number
violating processes with an associated large gauge boson multiplicity.

\item[] {\em{Method of solution: }}generation of phase
space configurations using
an efficient algorithm, coupled to a phenomenological
model of the physical
processes under study.

\item[] {\em{Typical running time: }}2-3 hours for a $10^4$ event
simulation (DEC ALPHA 4000), dependent on the process simulated

\item[] {\em{Unusual features of the program: }}none.

\end{itemize}

\newpage
\setcounter{page}{1}

\section{Introduction}

It has been known for some time that topologically non-trivial gauge
field configurations induce baryon- and lepton-number violating
({\BLNV}) vertices in the electroweak sector of the standard
model~\cite{th76a,th76b}. More recent investigations of these processes
suggest that such processes may be observable at supercollider energies.
An extensive review can be found in \cite{ma92}.

However, at present it is not possible to make definite predictions
even about the rate of such processes. What is possible is to take the
main features of the calculations and use these characteristics to form
a phenomenological model of \BLNV\ processes. The aim of such a model is
to predict possible experimental avenues that can be explored, with the
possibility of improved theoretical calculations in the future allowing
the model to be refined. One such model has been presented in
\cite{gi94a,gi94b}, and in section~2 we review this model.

In section~3 we describe the implementation of the model as a Monte Carlo
simulation, {\HERBVI}. This package simulates hard scattering processes
involving {\BLNV}, and has been designed to take into account the
very high particle multiplicities associated with \BLNV\ processes.
In addition, the formation of the extra baryons or antibaryons from
the \BLNV\ process is dealt with.

Finally, in section~4 we make some concluding remarks.
In an appendix we describe the use of the \HERBVI\ package by detailing
how it interfaces with the general Monte Carlo simulation program
{\HW}~\cite{hw88,hw92}.

\section{Simulating baryon number violation}

In this section, the model of \cite{gi94a,gi94b} is briefly outlined. One
of the general properties that has to be addressed is the parametrically
large multiplicity of gauge bosons produced
in \BLNV\ interactions~\cite{ri90,es90}.
Consequently, we discuss in some detail the method of~\cite{mambo}
for the efficient \MC\ generation of many-body phase space
configurations.

\subsection{General properties}

The general features of instanton-induced \BLNV\ taken into account by
\HERBVI\ are:
\begin{itemize}
\item{Quarks and leptons of all families attached to the \BLNV\ vertex,
with total change in baryon and lepton number
\begin{equation}
\De B = \De L = -3.
\end{equation} }
\item{A steeply rising cross section, the rise being cut off at some
energy due to unitarity constraints.}
\item{A parametrically large boson multiplicity, $n_B$, with a typical
value being $n_B = 30 \sim 1/\al_W$.}
\end{itemize}
The \BLNV\ process is assumed to be initiated by a valence
quark-quark collision. Thus the reaction is
\begin{equation}
\label{eq:anoproc}
q + q \rightarrow 7 \bar q + 3 \bar l
 + n_B W\left( Z\right) + n_H H .
\end{equation}
where the two incoming quarks and one outgoing antiquark are
first-generation, and three outgoing antiquarks of each of
the second and third generations are produced. The matrix
element is taken to be constant, so that the distributions
of produced particles are determined by phase space. Two options
for the gauge boson multiplicity $n_B$ are provided in the program:
either a fixed multiplicity, or a distribution modelled on
the leading-order \BLNV\ matrix element. Alternatively, a user-supplied
routine can be used.  The Higgs boson multiplicity $n_H$ is assumed to
be tied to that of gauge bosons, as discussed in sect.~3.5 below.

For the energy dependence, a simple step-function of the quark-quark
centre-of-mass collision energy $\sqrt{\shat}$ is assumed:
\begin{equation}
{\hat{\si}}\left( \sqrt{\shat}\right) = {\hat{\si}}_0 \th\left(
\sqrt{\shat} - \sqrt{{\hat{s}}_0}\right)
\end{equation}
The reasoning behind a model of this form has been discussed in detail
in~\cite{gi94a}. Again, a user-supplied subroutine may substituted
if desired.

The threshold energy parameter $\sqrt{{\hat{s}}_0}$ used in
modelling the cross section is expected to be of the order
of the characteristic energy of these processes, the sphaleron
mass~\cite{km84}, $M_{sp} = {\sqrt{6}\pi m_w/\al_w} \sim
18$ TeV. The appropriate value of the cross section parameter
${\hat{\si}}_0$ is a matter of great theoretical uncertainty.
However, if one wishes to investigate the properties of events
rather than their expected number, the magnitude of ${\hat{\si}}_0$
is not relevant. For reference, we note that the unitarity bound
for S-wave processes is given by
\begin{equation}
{\hat{\si}}_U\left(\shat\right) = \frac{16\pi}{\shat}
\end{equation}
which is of the order of tens of picobarns at the sphaleron energy
$M_{sp}$.

In addition to \BLNV\ events, \HERBVI\ is able to
generate the main expected background process, non-perturbative
$B$- and $L$-conserving multi-W production. The reaction is assumed
to be
\begin{equation}\label{multiW}
q_1 + q_2 \rightarrow q_3 + q_4 + n_B W\left( Z\right) + n_H H .
\end{equation}
This process is also expected to occur at large boson
multiplicities, $n_B\sim 1/\al_w$, with an energy scale $\sim
m_w/\al_w$. A detailed discussion, based on the argument that at high
energies electroweak theory has a large, approximately constant total
inelastic cross section, can be found in {\cite{ri91b}}.
Clearly, these events will be like \BLNV\ ones, but
without any fermion number violation. The predominance of boson over
fermion production means that the experimental signatures of the two
types of events will be similar. It is also expected that the
two processes will have similar threshold behaviour and cross
sections. In \HERBVI\ they are therefore modelled in the same
way, apart from the difference in fermion production.

\subsection{Phase space generation}

The main requirement of a \MC\ event generator for simulating \BLNV\
processes is to generate configurations in phase space for
high particle multiplicities at high efficiency. Therefore, careful
consideration of the method used to generate phase space configurations
is required.

One approach is to generate $n$-body phase space by factorising the
process into $n-1$ two-body decays. This is the method used in, for
example, the phase-space generator {\s{GENBOD}}~\cite{genbod}.
However, this method results in non-uniform phase space distributions,
and in particular becomes very inefficient at high multiplicities. In
{\cite{rambo}}, the efficiency of this method is quoted as being
less than 1\% for $n>17$.

The \MAMBO\ algorithm~\cite{mambo} is
designed to overcome this problem.
The aim is to generate efficiently $N$ momenta $p^\mu_i$ such that
the total momentum
\begin{equation}
P^\mu = \sum_{i=1}^{N} p^\mu_i
\end{equation}
is equal to the total energy and momentum required. In the rest frame of
the interaction, this is simply
\begin{equation}
P^0 = \sqrt{\shat},\ \ {\vec{P}} = 0 .
\end{equation}
In order to develop an efficient algorithm, one may divide the process
into two steps. Firstly, $N$ momenta $k_i^\mu$ are generated,
with the only constraint being the mass of each
particle, $k_i^2 = m_i^2$.
Obviously, their sum will not satisfy
the constraints. The second step is to transform these momenta, using
some combination of a Lorentz boost and scaling transformation, so that
the resultant momenta $p^\mu_i$ do satisfy the constraints. Note that
we may have different scaling transforms for the space-like and
time-like components of the momenta.

The key to the \MAMBO\ algorithm is the scaling transformion used. First,
assume that a suitable Lorentz transformation, acting on the $k^\mu_i$
to produce $q^\mu_i$, has been performed so that the system is now at
rest. Therefore
\begin{equation}
Q^\mu  =  \sum_{i=1}^{N} q^\mu_i
\end{equation}
such that ${\vec{Q}} = 0$. The scaling
transform is now performed by setting
\begin{equation}
q^\mu_i \rightarrow xq^\mu_i\;,\;\;\; x=\sqrt{\shat}/Q^0\;.
\end{equation}
Therefore, the total energy is now $\sqrt{\shat}$ but the particle masses
have been rescaled to become $xm_i$. The last step is to now find a \ze\
such that
\begin{equation}
\sum_{i=1}^{N}\left( \ze^2 {\vec{q}}_i^{\,2} + m_i^2\right)^{1/2} =
\sqrt{\shat} ,
\end{equation}
and then make the replacement
\begin{eqnarray}
{\vec{p}}_i & = & \ze {\vec{q}}_i \nonumber \\
p^0_i & = & \sqrt{\ze^2{\vec{q}}_i^{\,2} + m_i^2}.
\end{eqnarray}
The \MC\ weight associated with this procedure is
\begin{equation}
w = \ze^{3N-3}\prod_{i=1}^{N} \left( \frac{q^0_i}{p^0_i} \right)
\frac{\sqrt{\shat} -\sum_{i=1}^{N} x^2m_i^2 /q^0_i}{\sqrt{\shat}
-\sum_{i=1}^{N} m_i^2 /p^0_i} .
\end{equation}
It can be shown that $x\le 1$ implies $w\le 1$. Therefore, by
enforcing the condition $x\le 1$ and rejecting events with $w$ greater
than a uniform random number between $0$ and $1$, we can use this
algorithm to obtain an unweighted distribution.

The test results described in~\cite{mambo}\ indicate that, for
the typical configurations generated by {\HERBVI}, the efficiency is in
the region of 30\% for $n_B \sim 30$, falling to $\sim 6$\% at
the upper end of the range, $n_B \sim 120$.

\section{The HERBVI package}

The \HERBVI\ package is designed as an `add-on' hard process generator
for the \MC\ event generator {\HW}~{\cite{hw88,hw92}}. The latter
program is a general
purpose generator for high energy processes involving hadrons. In order
to explain how the two packages relate to each other, we now give a
quick review of the various stages of event generation in {\HW}.

The first stage is the generation of the hard process under
study. This can be either one of the processes already within {\HW}, or
alternatively a separate one such as those supplied by {\HERBVI}. The
kinematics and parton flavours are generated according to the matrix
elements under study, within appropriate phase space limits.

Next, the package simulates the perturbative evolution of the incoming
and outgoing partons, by generating parton showers. One of the features
of \HW\ is that it correctly treats \QCD\ coherence effects, which lead
to an ordering of the emission angles of the partons emitted in the
showers. This ordering results in a decrease in the
transverse momentum as one
moves away from the hard process towards either the initial state beam
particles or the final state products. The showering proceeds until
either the transverse momentum of emitted partons is below a cut off
(final state showers) or when the evolution scale is small enough to be
cut off by a scale associated with the structure function (initial state
showers).

The end of the showers represents the point where non-perturbative
effects lead to hadronisation. \HW\ uses a cluster
model~\cite{cluster} to
describe this process. Gluons in the parton
shower are split non-perturbatively
to $q{\overline{q}}$ pairs, and colour singlet pairs are joined together
to form colourless hadronic clusters. These clusters then decay into
hadrons, a process simulated using a simple phase space model. The decay
of unstable hadrons into stable products is then modelled using particle
data tables based on experimental measurements. Heavy quark decays are
treated by regarding them as new sources of coherent radiation and
repeating the process.

With this sketch of the \HW\ \MC\ generator in mind, we now turn to the
description of the \HERBVI\ package. Essentially, this models
\BLNV\ by generating the appropriate
hard process for {\HW}, although
the presence of baryon-number violation requires careful treatment
when hadronisation is modelled.  The package is described by
considering various routines within it, in an order corresponding
roughly to the order in which they are used by {\HW}.

\subsection{Initialisation}

\HERBVI\ contains an initialisation procedure designed to make the
changing of default values as easy as possible. Without any
modification, the package calls the initialisation routines at
the start of the generation of the first event.

\routine{HVHINI} - Initialise event generation variables

This routine performs the initialisation of the variables used in event
generation. It must be called if certain parameters are changed
by the user. The parameters available to the user are, at present,
the components of the \CKM\ matrix (stored in the array {\tt{HVCKM}}),
the cross section parameter {\tt{HVPLCS}}, and the limits on the
quark-quark subprocess energy-squared \shat.
The lower limit on $\shat$ corresponds to the threshold
energy of the model discussed in section 2.
The limits for \shat\ generation are set by
the variables {\tt{SLMIN}} and {\tt{SLMAX}}.
These variables are the logarithms of the ratios
{\tt{SLOWER}} $=\shat_{0}/{s}$ and
{\tt{SUPPER}} $=\shat_{max}/{s}$,
where $s$ is the total centre-of-mass energy squared.

\routine{HVINIT} - Main initialisation routine

This routine is only executed once per program run, and does nothing
if called again. All the parameters and flags are set to
default values within this routine.

Should the user wish to
change any default values at the start of the run, then a call should be
made to {\tt{HVINIT}} first, and the parameters then changed. A call
will then be made automatically to {\tt{HVHINI}} at the start of the
first event to recalculate derived parameters. If the user wishes to
change parameters during a run, another call to {\tt{HVHINI}} must
be made after the changes.

\subsection{Event generation}

The \MC\ program \HW\ uses the \HERBVI\ package for process codes
{\IPROC}$= 7000$ to $7999$. We now
examine in detail the various steps the
package performs for event generation in this case.

\routine{HVHBVI} - Main interface routine

This routine provides the main event generation interface to {\HW}.
Its purpose is to translate \HW\ variables into the appropriate
subroutine calls. This is achieved using the \HW\ variables
\IPROC\ and {\GENEV}. The first of these determines the type
of process to be generated. The values used in \HVN\ are
listed in Table~\ref{tab:iprocs}.
\begin{table}
\begin{center}
\begin{tabular}{|c|c|} \hline
\multicolumn{2}{|c|}{HERBVI process codes} \\ \hline \hline
{Code}&{Function} \\ \hline
7000-7099 & BLNV process \\
7100-7199 & Multi-W process \\
7200-7299 & Random selection \\ \hline
\end{tabular}
\end{center}
\caption{\HVN\ process codes. The random selection is of any
code in the range 7000 to 7199 with equal
probability.}\label{tab:iprocs}
\end{table}

The variable \GENEV\ determines the type of routine to be called. In
order to describe its use, we recall first how it functions within
the \HW\ package. In \HW, a hard process generation routine is
called twice. The first call, with \GENEV\ set to {\tt{.FALSE.}},
requests the generation of an event weight by the hard process routine.
The second call, with \GENEV\ now {\tt{.TRUE.}}, asks for event generation to
be completed and the event record to be written into the standard
\s{HEPDATA}\ common block. The \HERBVI\ package is structured in a
different manner, with separate routines for weight generation
({\tt{GENEV=.FALSE.}}), and writing the event data ({\tt{GENEV=.TRUE.}}).
The event parameters are passed between the two sets of routines
using common blocks.

The other function of {\tt{HVHBVI}} is a call to {\tt{HVHINI}}, in order
to initialise variables. This only occurs during the first execution of
the routine in each run.

Note that the dummy version of {\tt{HVHBVI}} must be deleted from the
\HW\ package to allow use of {\HERBVI}.

\subsection{Weight generation}

There are two routines to generate event weights, to cover the \BLNV\
and multi-W cases. We shall discuss the \BLNV\ case first, and then
highlight the differences for the $B$- and $L$-conserving routine.

\routine{HVHEVT} - \BLNV\ weight generation

This routine generates an event weight according to the \BLNV\ model.
The first function of the routine is to select the incoming partons,
and generate the centre-of-mass energy of the hard process using
{\tt{HVSGEN}}. Note that the only incoming partons allowed at
present are $u$ or $d$ quarks. This restriction is justified
since only collisions of protons and/or heavy ions are envisaged,
and the high threshold energies of \BLNV\ processes correspond
to a high momentum fraction $x$ even at supercollider energies.
Consequently valence quark distributions are the only significant
source of incoming partons.

The next distribution generated is that
of the bosons, accomplished using {\tt{HVRBOS}}. Two
multiplicity distributions are provided by this routine:
a constant or a leading-order type of distribution, selected
by means of the logical flag {\tt{HVCONT(7)}}.
These choices correspond to the two cases discussed
in~\cite{gi94a}. Alternatively, it is straightforward for
users to provide their own distributions. More details
are given in sect.~3.5 below.

The fermions produced in the hard process are now selected according to
a simple phase space model. One member of each left-handed doublet has
to be chosen. If the masses of the two particles are $m_1$ and $m_2$,
then the probability of choosing particle 1 instead of 2 is
\begin{equation}\label{eq:pest}
p = \left( 1+
\frac{m_2 K_1\left(m_2\be\right)}{m_1
K_1\left(m_1\be\right)}\right)^{-1}
\end{equation}
for massive particles, where \be\ is estimated using
\begin{equation}
\be = \frac{3}{2}\frac{n_B}{{\sqrt{\shat}} - n_B m_B},
\end{equation}
and $K_1$ is the modified Bessel function of the second kind.
For the case $m_1=0$, Eq.~(\ref{eq:pest}) has to be replaced by
\begin{equation}           \label{eq:pest2}
p = \left( 1 + {m_2\be K_1\left(m_2\be\right)} \right)^{-1}.
\end{equation}
A good approximation, used in {\HERBVI}, is to set $p=\myfrac{1}{2}$
except when selecting from ${\overline{t}}{\overline{b}}$ doublets.
The error caused by this approximation is negligible, provided
$n_B m_B/\sqrt{\shat}<\myfrac{1}{2}$.

As pointed out in~\cite{gi94a}, calculations of fermion number
violation by instanton mediated processes deal with gauge-eigenstate
down-type quarks, rather than the physical mass eigenstate particles.
Therefore, the outgoing quark distribution has to be projected through
the \CKM\ matrix. This is achieved by a call to {\tt{HVHPQM}}, provided
the control flag {\tt{HVCONT(3)}} is {\tt{.TRUE.}}.

Now that the fermion distribution has been fixed, charge conservation
for the whole process is enforced by a second call to {\tt{HVRBOS}}, and
the structure of the event stored for later use. The $x$ distributions
of the two incoming partons can now also be generated, by a call to
{\tt{HVSFUN}}.

The last step, if requested by changing the value of
the flag {\tt{HVCONT(1)}} to {\tt{.FALSE.}}, is to generate
the \MAMBO\ weight for the phase space distribution.
The anticipated normal use of the package is for an unweighted
distribution to be generated during the event data phase of
operation, but this can be overridden should the user wish to use a
weighted distribution. The main use we foresee for this option is the
generation of distributions corresponding to non-trivial matrix
elements.

\routine{HVHMWE} - Multi-W weight generation

The only fundamental difference between this routine and {\tt{HVHEVT}}
is in the nature and number of the outgoing fermions, since $B$ and $L$ are
conserved in this case. Again, only $u$ and $d$ quarks are allowed
as the incoming partons, $q_1$ and $q_2$ in Eq.~(\ref{multiW}),
and here the same applies to the outgoing partons, $q_3$ and $q_4$.

\subsection{Event data}

The event data insertion routines are executed if
the event weight has been accepted by {\HW}.
As for the weight generation phase, there are separate routines to cover
the two types of processes. Again, we describe the \BLNV\ case first and
then focus on the differences in the multi-W case.

\routine{HVHGEN} - Generate \BLNV\ event data

This routine writes the event data into the {\s{HEPDATA}} common block,
and sets \HW\ variables as appropriate. If the flag {\tt{HVCONT(1)}} is
{\tt{.TRUE.}}, its default value, then the \MAMBO\ algorithm is
used to generate an unweighted phase space configuration.

The \HW\ book-keeping is performed by a call to {\tt{HWEONE}}, which sets
up a $2\rightarrow 1$ subprocess. The internal \HW\ pointers are then
modified to take account of the final state particles. In particular,
the colour connection pointers for the fermions are set in such a way
that the outgoing light antiquark is connected to the incoming quarks,
and the three outgoing quarks of each higher generation are connected
to each other, so as to form three distinct colour singlets.

\routine{HVHMWG} - Generate multi-W event data

The main difference between this and {\tt{HVHGEN}} is the colour
structure of the final state. The colour connections to be made
between the partons are now those appropriate to quark-quark
elastic scattering. For this purpose, the routine {\tt{HVETWO}}
is used to set up a $2\rightarrow 2+n$ process.
This is a modification of the \HW\ routine {\tt{HWETWO}}, which
generates a $2\rightarrow 2$ process with the correct
colour pointers.

\subsection{Generation of distributions}

There are a number of routines within \HERBVI\ to generate the various
distributions. They are listed here together for reference. Note that
these routines obtain random numbers by calls to random
number generation routines within the \HW\ package (named
{\tt{HWR***}}).

\routine{HVHPQM} - Project quarks through \CKM\ matrix

The elements of the \CKM\ quark mixing matrix are used to project the
down-type quarks from their gauge eigenstates onto mass eigenstates.
Note that if for any reason the \CKM\ parameters are changed by the user,
the initialisation routine {\tt{HVHINI}} must be called to reset the
parameters used by this routine.

\function{HVRBIN} - Binomial distribution

Generation of the binomial distribution
\begin{equation}
\calP\left( n\right) = \frac{N!}{n!(N-n)!} p^n
\left( 1-p\right)^{N-n}.
\end{equation}
The distribution is generated by first generating two uniformly random
variables $u$ and $v$, such that $u^2 < \calP\left( v/u\right)$, and
then setting $n = ${\tt{INT}}$\left( v/u\right)$.

The main use of this
routine is in the generation of the $\ga /Z^{0}$ distribution by
splitting the number of neutral bosons $N$ into photons and $Z^0$'s with
probabilities $p=\sin^2 \th_w$ and $1-p=\cos^2 \th_w$ respectively.

\routine{HVRBOS} - Boson number distribution

This routine has two purposes, determined by a flag when
it is called. The first is to generate the total number
of gauge and Higgs bosons in the final state. The second function is to
ensure charge conservation by adjusting the relative numbers of charged
bosons. There are three control flags {\tt{HVCONT(7,8,9)}}
for this routine.

The flag {\tt{HVCONT(7)}} controls the number of gauge bosons, $n_B$. If
{\tt{.TRUE.}}, the default, then $n_B = 30$. Otherwise, the number is
selected according to a parametrisation of the multiplicity distribution.
The default parametrisation is of the form
\begin{equation}
\calP\left( n,\shat\right) = \exp \left( -\frac{\left(
n+a\right)^2}{b}\right)
\end{equation}
where
\begin{eqnarray}
a & = & {\mbox{\tt{-21.497D0}}} +
{\mbox{\tt{0.590D-2}}}\times \frac{\sqrt{\shat}}{\mathrm{GeV}}
\nonumber \\
b & = & {\mbox{\tt{0.483D-1}}} +
{\mbox{\tt{-0.854D-6}}}\times \frac{{\sqrt{\shat}}}{\mathrm{GeV}}
\end{eqnarray}
are parameters found by fitting to the results of phase space
calculations using the leading order matrix element for
instanton-induced {\BLNV}~\cite{fa90}.

The routine can be replaced by a user-supplied version, {\tt{HURBOS}},
by setting {\tt{HVCONT(8)}} to {\tt{.FALSE.}}. Note that both functions
of the routine have to be catered for, and that the dummy version of
{\tt{HURBOS}} must then be deleted from the \HERBVI\ package.

The final flag {\tt{HVCONT(9)}} controls the number of Higgs bosons
generated. The default is to generate according to a simple model,
namely that the probability of generating a Higgs boson is $1/16$ of
the probability of emitting a gauge boson from the \BLNV\ vertex.
This gives a reasonable agreement with the phase space
calculations performed in the energy region of interest with $m_H=300$
GeV. Note that the leading-order instanton calculation yields
\begin{eqnarray}
{\overline{n}}_B & \sim & \frac{3}{2}\frac{\pi}{\al_w} \left(
\frac{E}{M_{sp}}\right)^{4/3}, \\
{\overline{n}}_H & \sim & \frac{3}{32}\frac{\pi}{\al_w} \left(
\frac{E}{M_{sp}}\right)^{2}
\end{eqnarray}
which, for the region of interest $E \sim M_{sp}$, also predicts the
ratio $n_H/n_B \sim 1/16$.
The phenomenology of Higgs boson production in \BLNV\ interactions has not
yet been considered in detail, essentially because of this expected
strong suppression relative to gauge boson production.
Setting {\tt{HVCONT(9)=.FALSE.}} turns off Higgs boson generation
completely.

Once the charge distribution of the particles has been fixed, the
numbers of \ga\ and $Z^0$ bosons are determined using {\tt{HVRBIN}}
as outlined above.

\routine{HVSFUN} - Parton momentum fraction generation

This routine generates the momentum fractions $x$ of the incoming
partons. It requires the centre-of-mass energy of the hard subprocess to
be already determined. A weight is generated according to the parton
distribution functions of the colliding hadrons.

\routine{HVSGEN} - Hard process energy

Generation of the total centre-of-mass energy \shat\ for the hard
subprocess. The default is to generate \shat\ according to
the distribution $d\shat/\shat$. Setting the flag {\tt{HVCONT(4)}} to
{\tt{.FALSE.}} changes this distribution to $d\shat/\shat^2$. A
user-supplied routine, {\tt{HUSGEN}}, can be called instead of
{\tt{HVSGEN}} if desired by setting {\tt{HVCONT(6)}} to {\tt{.FALSE.}}.

Note that generation of a $d\shat/\shat$ distribution is equivalent to
postulating a {\em flat} parton-level cross section. This can be seen by
noting that the convolution of the parton level cross section with the
proton structure functions,
\begin{equation}
\si\left( s\right) = \int_{0}^{1} dx_1
\int_{0}^{1}{\mathrm{d}}x_2 \int_0^{s}
{\mathrm{d}}\shat f_1\left( x_1\right) f_2\left(
x_2\right) \de\left(\shat -x_1x_2s\right)
{\hat{\si}}\left(\shat\right) ,
\end{equation}
gives
\begin{equation}
\si\left( s\right) = \int_{0}^{1} \frac{{\mathrm{d}}\tau}{\tau}
{\hat{\si}}\left({\tau}s\right)
\int_{\tau}^{1} \frac{{\mathrm{d}}x}{x}  x f_1\left( x\right)
\frac{\tau}{x} f_2\left( \frac{\tau}{x}\right) ,
\end{equation}
where $\tau=\shat /s$.
Specialising to the case of a threshold parton-level cross section
\begin{equation}
{\hat{\si}}\left(\shat\right) = \hat\si_0 \th\left(
\sqrt{\shat}-\sqrt{{\hat{s}}_0}\right) ,
\end{equation}
we find
\begin{equation}
\si\left( s\right)  =  \hat\si_0 \int_{\tau_0}^{1}
\frac{{\mathrm{d}}\tau}{\tau} \int_{\tau}^{1}
\frac{{\mathrm{d}}x}{x}
x f_1\left( x\right) \frac{\tau}{x}f_2\left(\frac{\tau}{x}\right) ,
\end{equation}
where now $\tau_0 = \shat_0 /s$.
Therefore, to generate a flat parton-level cross section, we generate
$\tau$ according to $d\tau/\tau$ (corresponding to $d\shat/\shat$) and
$x$ according to $dx/x$, then reweight using the structure functions
$zf_i(z)$ at $z=x$ and $z=\tau/x$.

Recall that the range of $\shat$ is limited by the parameters
{\tt{SLMIN}} and {\tt{SLMAX}}, which are the logarithms of
{\tt{SLOWER}} and {\tt{SUPPER}} where
\begin{equation}
{\mbox{\tt{SLOWER}}} \le \frac{\shat}{s} \le {\mbox{\tt{SUPPER}}}\;.
\end{equation}
The default values correspond to the range
\begin{equation}\nonumber
0.24 \le \frac{\shat}{s} \le 0.8
\end{equation}
which gives $\sqrt{\shat}$ between approximately 50\% and 90\% of the total
energy. The upper limit is somewhat arbitrary; the rapid decrease
of the structure functions at large $x$ means that the contribution from
the upper end of the energy spectrum is negligible. This limit
has been introduced to increase the efficiency of the {\MC}.

The user can change the variables {\tt{SLMIN}} and {\tt{SLMAX}} to
alter the range of {\shat} generation at any time, although changing
these parameters in the middle of simulation requires a call to
{\tt{HVHINI}} to enable the derived parameters to be recalculated.

\subsection{Hadron formation}

The colour structure of \BLNV\ processes requires special treatment
during the later stages of event processing by {\HW}. As explained
above, there are three
colour-singlet three-quark (antiquark) vertices
associated with the process, the first of which
comprises the two incoming quarks and one outgoing antiquark.
The other two correspond to outgoing antiquark triplets.

These vertices do not occur
in any baryon-number conserving process,
and therefore must be treated by the \HERBVI\ package. However, the
set of three partons forming a singlet in the hard subprocess will not
necessarily be those forming the `extra' baryon as a result of {\BLNV},
because the first gluon emitted by any of these partons will carry
with it the colour connection to the \BLNV\ vertex. Therefore, the parton
cascade within \HW\ must be allowed to occur before the colour connections
are made. The point in \HW\ where this occurs is at
the start of the clustering phase of the
simulation, in the routine {\tt{HWCFOR}}.

\routine{HVCBVI} - Find unpaired partons after {\BLNV}

This routine is called at the start of the \HW\ routine {\tt{HWCFOR}},
if a \BLNV\ event has been generated. It searches through the event
record to locate all partons triplets connected with the \BLNV\ vertices.
The end product of the routine is to produce, for each vertex, a
quark and associated diquark (or antiquark and antidiquark,
as appropriate) that are colour connected.

\HW\ is at present
unable to incorporate heavy quarks into diquarks. This means that a
vertex such as ${\bar{b}}{\bar{b}}{\bar{u}}$ requires
special treatment. The approach used is to form a cluster from two
of the \BLNV\ particles, one of which is a heavy antiquark $\overline{Q}$.
The cluster is then split, producing an extra $q{\bar{q}}$ pair. A
mesonic colour singlet is formed from $q{\overline{Q}}$ leaving the light
antiquark $\bar{q}$ colour-connected to
the \BLNV\ vertex. This approach fails when
there is insufficient energy to form the extra $q{\bar{q}}$ pair.
In this case the event is rejected by calling the \HW\ error handler
{\tt{HWWARN}} with a negative error code, allowing the event
to be rejected in a controlled manner. Typically, about 1\%
of events have to be rejected in this way.

At the end of this routine, all the \BLNV\ colour connections have been
correctly made and the \HW\ package can now complete the process of event
generation. Note that, in order to
generate \BLNV\ events, the dummy
version of {\tt{HVCBVI}} must be deleted
from {\HW}.

\subsection{Analysis of events}

The \HERBVI\ package contains a routine to aid in the analysis of
\BLNV\ events. Whilst this code is not a necessary part of the package,
it does aid in the identification of simulated
particles direct from the \BLNV\
vertex as opposed to those formed as decay products of other unstable
particles.

\routine{HVANAL} - Analyse final state products

The event record is scanned to identify
the particles from the \BLNV\ vertex. Particles are flagged as `primary'
from the \BLNV\ vertex if they are either (anti)leptons from the vertex
itself, photons from the vertex, or (anti)baryons formed from the colour
triplet at the vertex. In addition, some other useful parameters such
as the rapidity, pseudo-rapidity and transverse momentum of each
particle are also calculated. The data are stored in the
common block {\tt{COMMON~HVADAT}}.

The contents
of {\tt{HVADAT}} are complementary to {\HW}\ output contained in the
standard {\tt{HEPEVT}} common block, and are listed
in Table~\ref{tab:comblk}. Note that the file {\tt{HERBVI10.INC}}
contains the variable declarations for {\tt{HVADAT}}. The
variable {\tt{FSPPTR}}, which is also stored in this common
block, contains the total number of final state particles found
and processed by {\tt{HVANAL}}.

\begin{table}
\begin{center}
\begin{tabular}{|l|c|} \hline
\multicolumn{2}{|c|}{{\tt{HVADAT}} contents} \\ \hline \hline
\multicolumn{1}{|c|}{Array} & {Contents} \\ \hline
{\tt{IDHNUM}} & Pointer to entry in {\tt{HEPEVT}} \\
{\tt{PSRFSP}} & Pseudo-rapidity of particle\\
{\tt{RAPFSP}} & Rapidity of particle\\
{\tt{PTFSP}} & Transverse momentum of particle\\
{\tt{IDHFSP}} & Type of particle (PDG code)\\
{\tt{ISFSP}} & {\tt{TRUE}} for particle from \BLNV\ vertex
\\ \hline
\end{tabular}\end{center}
\caption{Contents of the {\tt{HVADAT}} common block. Each
array listed has one entry for each particle
in the final state after the \HW\ simulation phase.}
\label{tab:comblk}
\end{table}

\subsection{Program summary}

For convenience we list here the main routines, control flags and default
settings. The
routines are listed in Table~\ref{tab:rtns}. Other auxiliary routines,
not listed here, all have names starting with either {\tt{HV****}} or
{\tt{BVP***}}. This convention has been followed to avoid clashing with
other software. These routines are associated with the internal workings
of the \HERBVI\ package and do not concern us here.

\begin{table}
\begin{center}
\begin{tabular}{|c|c|} \hline
\multicolumn{2}{|c|}{HERBVI routine names} \\ \hline \hline
{Routine}&{Function} \\ \hline
{\tt{HVANAL}} & Initial analysis\\ \hline
{\tt{HVCBVI}} & Final state clustering \\
{\tt{HVCBVT}} & Find hard parton parent \\ \hline
{\tt{HVETWO}} & $2\rightarrow 2+n$ hard process \\ \hline
{\tt{HVHBVI}} & \HW\ interface routine \\
{\tt{HVHEVT}} & \BLNV\ weight generation \\
{\tt{HVHGEN}} & \BLNV\ event data \\
{\tt{HVHINI}} & Initialise derived parameters \\
{\tt{HVHMWE}} & Multi-W weight generation \\
{\tt{HVHMWG}} & Multi-W event data \\
{\tt{HVHPQM}} & \CKM\ matrix projection \\ \hline
{\tt{HVINIT}} & Main initialisation \\ \hline
{\tt{HVRBIN}} & Generate binomial distribution \\
{\tt{HVRBOS}} & Generate boson distribution \\ \hline
{\tt{HVSFUN}} & Generate parton $x$ distribution \\
{\tt{HVSGEN}} & Generate \shat\ distribution \\ \hline
\end{tabular}
\end{center}
\caption{Main \HERBVI\ routines.}\label{tab:rtns}
\end{table}

The control flags are listed in Table~\ref{tab:cflags}. The default
value in all cases, as set by {\tt{HVINIT}}, is {\tt{.TRUE.}}.
The default values of physical parameters are in general taken from the
corresponding \HW\ ones. The only exception to this are the limits
for the $\shat$ generation, which are given default values by
{\HERBVI} as described in section 3.5.

\begin{table}
\begin{center}
\begin{tabular}{|c|c|c|c|} \hline
\multicolumn{4}{|c|}{HERBVI control flags} \\ \hline \hline
{Flag}&{\tt{.TRUE.}}&{\tt{.FALSE.}}&{Comments}\\ \hline
{\tt{HVCONT(1)}}& Unweighted & User weighted & Phase space generation \\
{\tt{HVCONT(2)}}& \BLNV & Multi-W & (internal flag) \\
{\tt{HVCONT(3)}}& On & Off & \CKM\ matrix projection \\
{\tt{HVCONT(4)}}& On & Off & Uniform \shat\ generation \\
{\tt{HVCONT(5)}}& Off & On & Cut-off on \shat\ generation \\
{\tt{HVCONT(6)}}& Off & On & User-supplied \shat\ generation \\
{\tt{HVCONT(7)}}& On & Off & Set $n_B=30$ \\
{\tt{HVCONT(8)}}& Off & On & User-supplied $n_B$ generation \\
{\tt{HVCONT(9)}}& On & Off & Allow Higgs production  \\ \hline
\end{tabular}
\end{center}
\caption{Control flags. The default setting for each flag is
{\tt{.TRUE.}}, as set by {\tt{HVINIT}}.}\label{tab:cflags}
\end{table}

\section{Conclusion}

We have described the various elements of
the \HERBVI\ program. This package takes the form of
a hard scattering process event generator that interfaces to
\HW, which is a general-purpose Monte Carlo program for
simulating particle interactions involving hadrons.

The phenomenological model which forms the motivation behind the
\HERBVI\ package is based upon the features of instanton-induced
\BLNV\ calculations, and we have briefly reviewed its features.
Of particular importance from an event generation point of view
is the parametrically large number of gauge bosons associated with
a \BLNV\ event, and we have addressed the issue of efficient phase
space generation for such outgoing final states. The
simulation of high-multiplicity gauge boson production with
conservation of baryon- and lepton- number, which is expected to
be the principal background to the \BLNV\ processes considered
here, is also implemented.

The violation of baryon number requires particular attention
when interfacing to a general simulation package such
as {\HW}. Specifically, the formation of the
extra (anti)baryons that violate baryon number
conservation requires the correct colour connections
between partons to be made after the emission of
\QCD\ radiation from the partons involved in the hard scattering.
The interface between the \HERBVI\ package and \HW\ allows these
connections to be made at the appropriate time during event generation.

Finally, we have described the provision that has been made within
\HERBVI\ to allow users to provide their own distributions. The
distributions contained within the package correspond to the models
discussed in~\cite{gi94a}. The interface to user-supplied
distributions has been constructed in a straightforward manner to
enable the study of \BLNV\ processes with other parton-level
cross sections and boson multiplicity distributions.

\section*{Acknowledgement}

We are most grateful to Andreas Ringwald for many helpful
discussions and suggestions during the development of the program.

\section*{Appendix A: Installing HERBVI}

The \HERBVI\ package is designed to be used with the \HW\ Monte Carlo
simulation package. Versions 5.7 and later of \HW\ contain hooks
that allow the two packages to link together in a relatively
simple manner. There are two steps required to link the packages.

\begin{itemize}
\item The dummy subroutines {\tt{HVCBVI}} and {\tt{HVHBVI}} must be
deleted from the \HW\ package. Failure to do this will cause the
program to terminate with an error message when the first attempt
to generate a \BLNV\ event is made.

\item The parameters {\tt{NMXHEP}}, {\tt{NMXPAR}}, {\tt{MODMAX}} and
{\tt{NMXJET}} must be increased from their default values. The
default values are set in the file {\tt{HERWIGxy.INC}}, where
{\tt{x.y}} is the program version number, and the
(recommended) new values are
shown in Table~\ref{tab:newds}. If this change is not made
then \HW\ will run out of internal working space with unpredictable
results.
\end{itemize}

\begin{table}\begin{center}\begin{tabular}{|c|r|r|} \hline
\multicolumn{3}{|c|}{Parameter settings for \HERBVI} \\
\hline \hline
\multicolumn{1}{|c|}{Parameter}&\multicolumn{1}{|c|}{Default}&
\multicolumn{1}{|c|}{\tt{HERBVI}}\\ \hline
{\tt{NMXHEP}} & 2000 & 15000 \\
{\tt{NMXPAR}} & 500 & 2000 \\
{\tt{MODMAX}} & 5 & 200 \\
{\tt{NMXJET}} & 200 & 500 \\ \hline
\end{tabular}\end{center}\caption{Recommended \HW\ parameter settings
for use with the \HERBVI\ package.}\label{tab:newds}
\end{table}

The first change can be made by deleting the
dummy routines from the \HW\ source code.
The second change must be made
by editing the parameter settings
in the {\tt{HERWIGxy.INC}} file.
The changes in these parameters are required to
give the \HW\ package enough working space to process the
high-multiplicity \BLNV\ events. Note that the formatting of
printed output from \HW\ is automatically changed for
{\tt{NMXHEP}} values greater than 9999, in order to ensure that the
80-column output is readable.

It should be stressed that, after the above changes have been
made, the \HW\ package must be recompiled to reflect the change in
size of the common blocks.

Finally, we remind the reader that to use \HERBVI\ with user-defined
distributions ({\tt{HVCONT(6)}} and/or {\tt{HVCONT(8)=.FALSE.}}),
the appropriate subroutines {\tt{HUSGEN}} and/or
{\tt{HURBOS}} must be supplied by the user and the corresponding
dummy subroutines deleted from the {\tt{HERBVI}} package.

\section*{Appendix B: Test run output}

The output of the \HERBVI\ package, after processing by the
\HW\ program, is contained as a list of final state hadrons and leptons
within the standard {\tt{HEPEVT}} common block. As there are typically
$\sim 10^{3-4}$ particles produced per \HERBVI\ event, is is not
practical to show one of these events here. Instead, we show the output
of the test program supplied with the \HERBVI\ package. This program
generates 1000 events, with a total collider energy of 40~TeV, using
the {\tt{IPROC=7200}} option.

The implementation of random number generators is, generally
speaking, not machine-independent due to numerical precision effects.
Therefore, one should expect to get similar, but not necessarily
identical, results when comparing the output of \HERBVI\ runs from
different computer architectures.

\begin{verbatim}
          OUTPUT ON ELEMENTARY PROCESS

          NUMBER OF EVENTS   =         992
          NUMBER OF WEIGHTS  =       52930
          MEAN VALUE OF WGT  =  1.6247E-04
          RMS SPREAD IN WGT  =  6.0487E-04
          ACTUAL MAX WEIGHT  =  7.7377E-03
          ASSUMED MAX WEIGHT =  8.2604E-03

          PROCESS CODE IPROC =        7200
          CROSS SECTION (PB) =  0.1625
          ERROR IN C-S  (PB) =  2.6291E-03
          EFFICIENCY PERCENT =   1.967

 HERBVI parameters
 =======================================================
 Total collider centre-of-mass energy (GeV): 0.40000E+05
 Minimum root s value                 (GeV): 0.19596E+05
 Maximum root s value                 (GeV): 0.35777E+05
 Parton-level cross section            (nb): 0.10000E+01
 Number of gauge bosons set to fixed value :     30
 Distribution of Higgs boson number        :   Default

 Number of HVCBVI errors                   :       8
 Number of events                          :    1000
 Percentage of events with BVI rej.        :   0.800
 Number of BVI events                      :     502
 Number of Multi-W events                  :     498

\end{verbatim}

\end{document}